\documentclass{aa}
\usepackage{graphics,epsfig,lscape}
\usepackage{natbib}

\bibpunct{(}{)}{;}{a}{}{,}

\def\cm-2{cm$^{-2}$}

\def\chandra{{\it Chandra}}

\def\xmm{{XMM-Newton}}
\def\n253{\object{NGC~253}}
\def\m33{\object{M~33}}
\def\mx7{\object{M~33~X$-$7}}
\def\x7{\hbox{X$-$7}}

\begin{document}

\title{Spectroscopy of the brightest optical counterparts of X-ray sources in the direction of
   M~31 and M~33}

   \author{D.~Hatzidimitriou\inst{1} \and
       W.~Pietsch\inst{2} \and
       Z.~Misanovic\inst{2} \and
       P.~Reig\inst{3} \and
       F.~Haberl\inst{2}}

       \institute{University of Crete, Department of Physics,P.O.Box 2208, 71003,
       Heraklion, Greece \and Max-Planck-Institut f\"ur extraterrestrische Physik,
           85741 Garching, Germany \and IESL, Foundation for Research and Technology, 71110,
            Heraklion, Greece}

     \offprints{D.~Hatzidimitriou, e-mail: {\tt dh@physics.uoc.gr}}

   \date{Received; accepted }

    \abstract
     {Recent surveys of the Local Group spiral Galaxies M~31
     and M~33 with XMM-Newton yielded a large number of
     X-ray sources.}
     {As part of the effort to identify and classify the
     objects responsible for this X-ray emission, we have
     obtained optical spectra of the brightest optical
     counterparts of the identified X-ray sources, using the 1.3m
     Skinakas Telescope. Most of these objects are foreground star
     candidates. The purpose of the present study is to confirm this
     identification and to explore the compatibility between the optical spectral
     classification and the observed X-ray properties of the sources.}
    {We have obtained optical spectra for the 14 brightest optical
    counterparts of X-ray sources identified by XMM-Newton in the direction
    of M~31 and for  21 optical counterparts in the direction of M~33, using the 1.3m
    Skinakas telescope in Crete, Greece.}
    {All of the M~31 sources and all but one of the M~33
    sources were confirmed to be foreground stars, of spectral types between A and
    M.One of the stars is a late M dwarf with H$_{\alpha}$ emission, a flare star,
    also displaying strong X-ray variability. One of the M~33 sources (lying within
    the D25 ellipse) corresponds to a previously known background galaxy, LEDA 5899.}{}


   \authorrunning{Optical counterparts of X-ray sources in M31 and M33 field}

   \maketitle

\section{Introduction}
Recent surveys of the Local Group spiral Galaxies M~31 (Pietsch et
al. 2005a, hereafter PFH2005) and M~33 (Pietsch et al. 2004,
hereafter PMH2004) with XMM-Newton yielded a large number of X-ray
sources. With their moderate Galactic foreground absorption, both
galaxies are ideal for studying the X-ray source population and
diffuse emission in nearby spiral galaxies.

The Andromeda galaxy (M~31) is a massive SA(s)b galaxy, located at
a distance of 780 kpc (Holland 1998) and seen under an inclination
of 78$^o$ (de Vaucouleurs et al. 1991).

As described in detail in PFH2005, a total of 856 X-ray sources
were detected in an area of 1.24 square degrees. Within M~31, 21
supernova remnants (SNR) and 23 SNR candidates were detected, as
well as 18 super-soft source candidates, 7 X-ray binaries and 9
X-ray binary candidates, 27 globular cluster sources and 10
globular cluster source candidates, which most likely are low mass
X-ray binaries within the cluster.  From 567 hard sources, some
are expected to be X-ray binaries or Crab­like SNRs within M~31,
and the rest, background AGN. As well as the sources within M~31,
there were also foreground and background sources: six sources
were identified as foreground stars and 90 as foreground star
candidates, one as a BL Lac type active galactic nucleus (AGN) and
36 as AGN candidates. One source coincided with the Local Group
galaxy M~32, one with a background galaxy cluster and another is a
background galaxy cluster candidate.

M~33, an Sc galaxy, is located at a distance of 795 kpc (van den
Bergh 1991) and is seen under a relatively low inclination of
56$^o$ (Zaritsky et al. 1989). PMH2004 detected a total of 408
sources in a 0.8 square degree field combining the counts of all
EPIC instruments. About half of the sources lie within M~33; the
rest are background AGNs or foreground stars. Cross-correlation
with archival data led to the identification of 5 foreground stars
and 30 foreground star candidates,  as well as 12 AGN candidates.
Within M~33, 21 supernova remnants (SNR) and 23 SNR candidates
were identified, as well as 5 super-soft sources and 2 X-ray
binaries. From the remaining 267 X-ray sources classified as
"hard", those within M~33 may be low or high mass X-ray binaries,
or Crab-like SNRs, while a significant fraction of them are
expected to be background AGN. Misanovic et al. (2005) analyzed
the individual observations of the XMM-Newton data of M~33,
detecting 39 new sources and improving the positions of another
311. They also studied the variability of the sources on time
scales of hours to months or years, their spectral characteristics
and classification. For example, using the detected variability,
they were able to classify 8 new X-ray binary candidates.

Pietsch et al. (2005b) searched for X-ray counterparts of optical
novae detected in M~31 and M~33 and discovered  21 X-ray
counterparts for novae in M~31 - mostly identified as supersoft
sources by their hardness ratios - and two in M~33.

As part of the effort of identifying and classifying the objects
responsible for the observed X-ray emission in the direction of
M~31 and M~33, we have obtained optical spectra of the brightest
optical counterparts of the identified X-ray sources, using the
1.3m Skinakas Telescope. These brightest optical counterparts were
taken from PFH2005 for M~31 and from PMH2004 for M~33. Most of
these objects are foreground star candidates, as also mentioned in
the corresponding reference papers. The purpose of the present
study is to confirm this and to explore the compatibility between
the optical spectral classification and the observed X-ray
properties of the sources. One of the bright optical counterparts
that we have observed is the known galaxy LEDA 5899, which was
also part of the CFA redshift survey (Huchra et al. 1999).

In the next section, we describe the optical observations; in
Section 3 we describe the data reduction. In Section 4, the method
followed for the classification of the spectra obtained is
presented and in Section 5 we discuss the results, including
individual objects of interest.

\section{Optical observations}
The optical observations used in this study were carried out
during three observing runs, on September 6, 2003, on October
18-22, 2003 and on October 10-11, 2004, using the 1.3-m
Ritchey-Cretien telescope at Skinakas Observatory, located on the
island of Crete (Greece).

The telescope was equipped with a 2000x800 ISA SITe CCD camera and
a 1302 lines/mm grating, giving a nominal dispersion of 1.04
\AA/pixel, and a wavelength coverage from 4748 \AA~~ to 6828 \AA.
The spectral region was selected so as to include both
H$_{\alpha}$ and H$_{\beta}$ spectral lines. For the three spectra
that were obtained in September 2003, the wavelength region was
somewhat different, shifted further to the blue (from 3900 \AA~~
to 6000 \AA). One of these three spectra (of source number 3 in
M~33) was also observed in October 2003, with the wavelength
coverage used for the bulk of the spectra. As will be discussed
later, the two spectra obtained for this object served as a test
of the compatibility between the spectral classes obtained in the
two different wavelength regions.

Exposure times ranged from 600 to 7200~s, depending on the object's
magnitude and on seeing and weather conditions. In most cases, two
or more exposures were obtained per object. Each object
observation was followed by an arc calibration exposure (CuAr).

A total of 18 standards, of spectral types ranging from O9.5 to
M6, were also observed with the same configuration as the targets.

In the first observing run  three objects in the direction of M~33
were observed (6/9/2003). In the second run another 19 targets in
the direction of M~33 were observed (one of them repeated from the
first run, so a total of 21 targets in the direction of M~33 were
observed), while in the third, spectra for 14 objects in the
direction of M~31 were obtained. The first 6 columns of Table 1
list the details of the observations.

\begin{enumerate}
\item {\em Column 1} gives the X-ray source number of the target
as it appears in the corresponding X-ray catalogue paper, i.e., in
PFH2005 for M~31 and in PMH2004 for M~33. There is one exception,
object 3b in the direction of M~33, which was not detected by the
automatic procedures due to its far off axis location and thus it
was not included in PMH2004 (see Section 5.2 for more details).

For object 122 in the direction of M~33, the bright USNO-B1 star
1210-0020531 given as possible optical counterpart in PMH2004 is
actually resolved into two stars on deeper 4m Mosaic images of
M~33, separated by $\sim$6\arcsec\ in the NNE/SSW direction (see
Section 5.3 for more details). Spectra for both of these stars
were obtained.

If available, Chandra source names are given in the footnotes of
the Table. For M~31 only one source ([PFH2005] 31) has a Chandra
identification, while for M~33 two sources ([PMH2004] 196 and 200)
have also been identified in existing Chandra catalogues.

 \item {\em Columns 2\& 3} give the co-ordinates of the objects (epoch 2000).

\item {\em Column 4} provides the USNO-B1 identification derived
from the cross-correlation between the X-ray catalogue and the
USNO-B1 catalogue.

\item {\em Columns 5 \& 6} list the B and R magnitudes of the
corresponding USNO object (B2 and R2 in the USNO-B1 catalogue).
The limiting magnitude of the sample of stars observed in M~33 was
R$=$14.75, while in M~31 it was 12.43. In neither case is the
sample of objects observed complete to the corresponding limiting
magnitude. The brighter limiting magnitude for the M~31 stars was
dictated by observing time limitations only (as three of the
nights allocated were lost to poor weather conditions).

\item {\em Column 7} gives the 2MASS identification derived from
the cross-correlation between the X-ray catalogue and the 2MASS
catalogue.

\item {\em Column 8} provides the corresponding J magnitude of the
2MASS counterpart (as given in the 2MASS catalogue).

\item {\em Column 9} gives the date of the Skinakas observation.

\item {\em Column 10} lists the total exposure time for the
specific spectrum, while the number of exposures taken for that
object is provided in the parenthesis following the total exposure
time.
\end{enumerate}

Figure 1 illustrates how well the X-ray positions correlate with
the coordinates of the optical counterparts observed.

\begin{figure}
\caption[]{X-ray positions (3-sigma error circles in black) of
PMH2004 sources overlaid on the optical image of Massey et al.
(2002).The bright stellar counterparts with spectroscopic
follow-up of [PMH2004] 297 and 281 are clearly visible. The
smaller white circles show the improved positions and errors from
the analysis of individual XMM-Newton observations (Misanovic et
al. 2005). Note that object 281 was not detected in this latter
study. The figure also shows one additional X-ray source in the
FoV that was classified as 'hard' ([PMH2004] 304).}
\end{figure}

\begin{table*}
\begin{center}
\caption{Log of objects observed in the direction of M~31 and M~33
and derived spectral types.  }

\begin{tabular}{lcccrrcrccl}
\hline\hline\noalign{\smallskip}
\multicolumn{11}{c}{{M~31}}\\
\hline\noalign{\smallskip} {ID}$^*$  &  {RA}   & {Dec} &
\multicolumn{3}{c} {USNO-B1}  & \multicolumn{2}{c}{2MASS} & {Date}
& {Exp.} & {Spectral}
\\
      &(2000)&(2000)&{ID}&{B}&{R}&{ID}&{J (mag)}&&{(s)}&{type}\\
\noalign{\smallskip} \hline\noalign{\smallskip}
26  & 00 39 43.54 & 40 39 42.1 & 1306-0011237&13.00 &12.43&00394351+4039425&11.57& 11/10/04& 3600  (2) & G0 \\
31$^1$ &00 39 56.53& 40 41 00.2 & 1306-0011322&10.76 &10.31&00395652+4041003& 9.78& 10/10/04& 1200 (2) & F0 \\
49  & 00 40 13.79 & 40 35 33.9 & 1305-0011683&11.74 &11.46&00401383+4035322&10.69& 11/10/04& 2400 (2)& F5 \\
59  & 00 40 23.75 & 40 53 06.3 & 1308-0012001&10.84 &10.21&00402380+4053069& 9.70& 10/10/04& 1200 (2) & F5 \\
101 & 00 40 57.11 & 40 56 38.1 & 1309-0012722&12.00 &11.42&00405703+4056384&10.81& 11/10/04& 2400 (2) & F5 \\
137 & 00 41 24.35 & 40 55 34.6 & 1309-0012839&12.34 &11.49&00412410+4055333&12.09& 11/10/04& 2400 (2) & G8\\
168 & 00 41 43.39 & 41 05 06.0 & 1310-0013030&11.91 &11.17&00414345+4105047&10.35& 11/10/04& 2400 (2) & G0 \\
217 & 00 42 08.89 & 41 23 31.3 & 1313-0012724&11.06 & 9.78&00420901+4123306& 8.38& 10/10/04& 1200 (2) & G8 \\
464 & 00 43 32.68 & 41 09 11.1 & 1311-0013521&11.58 & 9.96&00433260+4109092& 8.62& 10/10/04& 1800 (2) & K0 \\
479 & 00 43 41.46 & 41 42 24.0 & 1317-0014537&11.24 &10.27&00433892+4138472&15.58& 10/10/04& 1200 (2) & G8 \\
498 & 00 43 50.31 & 41 24 11.4 & 1314-0013364&11.48 &10.92&00435017+4124115&10.58& 10/10/04& 2400 (2) & F7 \\
553 & 00 44 23.91 & 42 00 10.1 & 1320-0014047&11.51 &10.70&00442392+4200095&10.21& 10/10/04& 1200 (1) & G0 \\
615 & 00 45 07.61 & 41 53 57.2 & 1318-0015195&12.64 &11.95&00450752+4153581&10.22& 11/10/04& 2400 (2) & G9 \\
733 & 00 46 19.32 & 42 21 29.4 & 1323-0017450&12.20 &12.16&00461905+4221308&10.53& 11/10/04& 2400 (2) & G0 \\
\noalign{\smallskip} \hline\noalign{\smallskip}
\noalign{\smallskip}
\multicolumn{11}{c}{{M~33}}\\
\noalign{\smallskip}
\hline\noalign{\smallskip}
\noalign{\smallskip}
 3$^2$  &  01 31 54.90  & 30 29 54.5 &1204-0019506& 12.37&11.34&01315500+3029522& 9.73& 06/09/03& 2700 (2)& G9\\
   &&&&&&&& 18/10/03&600 (1)&\\
 3b$^3$  &  01 31 52.62  & 30 29 26.2 & 1204-0019500&17.46 & 14.66   &                &     & 18/10/03& 2400 (2) & M3  \\
28 &  01 32 23.37  & 30 47 48.8 & 1207-0019732&14.24 &11.65&                &     & 20/10/03& 1800 (3)& G9  \\
39$^4$ &  01 32 30.25  & 30 36 17.3 & 1206-0019336&16.61 &14.75&01323043+3036186&13.32& 20/10/03& 4800 (3) & K3:: \\
54 &  01 32 38.67  & 30 12 19.2 & 1202-0019688&13.40&10.99&01323878+3012167&11.34& 20/10/03& 1200 (2) & G4   \\
77 &  01 32 51.27  & 30 08 13.2 & 1201-0020193&12.55 &10.89&01325128+3008140&11.23& 20/10/03& 1800 (3)& F5  \\
92 &  01 32 56.75  & 30 15 44.2 & 1202-0019784&10.85 & 9.50&01325685+3015429& 8.20& 20/10/03&  600 (1) & G7 \\
   &               &            &             & &     &                &     & 21/10/03& 1200 (2)&      \\
122$^5$&  01 33 13.12  & 31 02 52.1 & 1210-0020531&15.61 &13.84&01331353+3102527&14.58& &  &    \\
   &  01 33 13.41  & 31 02 48.9 &              &  &   &                &     & 21/10/03& 1800 (1) & G0   \\
   &  01 33 13.63   & 31 02 52.9   &              &   &  &                &     & 21/10/03& 1800 (1) & G5:: \\
142&  01 33 22.95  & 30 56 55.0 & 1209-0020773&9.45 &9.04 &01332270+3056573&14.93& 22/10/03&  600 (2) & F4    \\
182&  01 33 37.04  & 30 23 23.2 & 1203-0021390&8.20 &7.96 &01333708+3023213& 7.58& 06/09/03&  480 (2) & A5  \\
196$^6$& 01 33 41.80 & 30 38 48.9 &           &   &    &01334186+3038491&12.61& 21/10/03& 7200 (4) & M5\\
200$^7$&  01 33 43.26  & 30 46 30.7 & 1207-0020534&11.58 &11.13&01334337+3046307& 9.77& 06/09/03& 1620 (2) & G5 \\
204&  01 33 46.51  & 30 54 32.3 & 1209-0020967&11.08&10.46&01334657+3054308& 9.91& 21/10/03& 1800  (1) & F4  \\
   &               &            &             & &     &                &     & 22/10/03& 3600 (2) &     \\
206&  01 33 46.66  & 30 54 53.0 & 1209-0020969&14.45 &12.21&01334671+3054550&11.93& 21/10/03& 1800 (1) & K0  \\
   &               &            &             & &     &                &     & 22/10/03& 3600 (2) &     \\
281&  01 34 26.34  & 30 58 02.7 & 1209-0021391&11.15 &10.48&01342664+3058028& 9.56& 22/10/03& 1200 (2) & F8  \\
297&  01 34 32.98  & 30 57 54.3 & 1209-0021478&13.91 &12.28&01343309+3057538&11.89& 22/10/03& 2400 (2) & G8\\
337&  01 34 52.99  & 30 28 12.0 & 1204-0021198&13.25 &11.92&                &     & 22/10/03& 1200 (2) & F5     \\
358&  01 35 08.29  & 31 02 18.7 & 1210-0021425&11.31 & 9.99&01350881+3102189&13.48& 18/10/03& 4500 (3) & galaxy \\
372&  01 35 15.42  & 30 55 09.0 & 1209-0021845&11.78 &10.98&01351564+3055084&10.53& 22/10/03& 1200 (2) & F8 \\
406&  01 35 51.66  & 30 44 52.3 & 1207-0021551&10.04 & 9.38&01355164+3044531& 8.66& 22/10/03&  600 (1) & F5  \\
\noalign{\smallskip} \hline\noalign{\smallskip}
\noalign{\smallskip}
\end{tabular}
\end{center}
$^*$ X-ray source numbers from PFH2005 (M~31) and PMH2004 (M~33)\\
$^1$ Chandra source s1-74, from Williams et al. 2004\\
$^2$ The adopted spectral type for this object is the average of
the spectral types obtained from the two observing runs.\\
$^3$ This object is a newly discovered source and it was not
included in the PMH2004 catalogue. See section 5.2 for details.
$^4$ far off-axis in Chandra observations 1730, not in
source list of Grimm et al. (2005)\\
$^5$ As explained in Section 5.3,  the bright USNO-B1 star
1210-0020531 is resolved into two stars  on  deeper 4m Mosaic
images of M~33, separated by $\sim$6\arcsec\ in NNE/SSW direction.
The coordinates of these two stars and the corresponding spectral
types we derived for them, are given in the two lines following
the entry for source 122.\\
$^6$ Chandra source CXO~J013341.8+303848, from Grimm et al. 2005;
    no USNO-B1 position, 2MASS position given.\\
$^7$ Chandra source CXO~J013343.4+304630, from Grimm et al. 2005\\
\end{table*}

\section{Data reduction}
The data reduction was performed using the {\em STARLINK} Figaro
package (Shortridge et al. 2001). The frames were bias subtracted,
flat fielded and corrected for cosmic ray events. The 2-D spectra
 were subsequently  sky subtracted using the {\em
POLYSKY} command. A spatial profile was then determined for each
2-D spectrum, and the  object spectra were optimally extracted
using the algorithm of Horne (1986), with the {\em OPTEXTRACT}
routine. Arc spectra were then extracted from the arc exposures,
using exactly the same profiles as for the corresponding object
spectra. The arc spectra were subsequently used to calibrate the
object spectra. After calibration, spectra from individual
exposures of the same object were co-added to yield the final
spectrum.

This procedure was followed for almost all of the observations,
with one exception, namely object USNO-B1 1210-0020531 in the M~33
list. In this case, a second object  was located very close to the
target (see section 5.2). The resulting spectra could not be
extracted separately using the optimal extraction procedure
described above. Due to the slight curvature of the spectra,
single linear extraction along the lines of the CCD was not
possible either. Thus we have used a step extraction with
sufficient wavelength overlap between subsequent extractions to
ensure compatibility of the fluxes. The calibrating arc spectra
were created in exactly the same way.

 The signal-to-noise ratio (S/N) of the final spectra varied from 20
 to about 350, with an average of 150.

The same reduction and calibration procedures were followed for
the 18 spectroscopic standard stars. Most of the standard spectra
have a S/N ranging between 200 and 350, while two have somewhat
lower S/N (around 130).

 Figure 2 shows six examples of target spectra (black lines), all of foreground stars
 of different spectral types, that have been flux-normalized for presentation
 purposes. For comparison, we show in grey the corresponding standard
 star spectrum (of the same spectral type) shifted in y by an arbitrary
 amount.

\begin{figure*}
\caption[]{Six examples of spectra (flux normalized) of optical
counterparts of X-ray sources in the direction of M~31 and M~33
(black lines). For comparison, we show in grey the corresponding
standard star spectrum (of the same spectral type) shifted in y by
an arbitrary. The identified lines are
 H$_{\alpha}$ (6563\AA), H$_{\beta}$ (4861\AA), FeI (5270, 5328,
 6495\AA) and NaI (5890\AA).}
\end{figure*}

\section{Spectral classification}
Classification of the obtained spectra was achieved via
cross-correlation with the 18 standard star spectra.

Each object spectrum was cross-correlated with each standard
spectrum, and the height of the corresponding cross-correlation
peak (hereafter, ccp) was recorded. These ccp heights were plotted
as a function of the spectral type of the standard, with the
maximum of the curve yielding the adopted spectral type for the
object spectrum. Figure 3 shows a typical example of such a plot,
for object USNO-B1 1201-0020193. The highest ccp is about 0.9,
corresponding to an F5 spectral type. For the entire sample, the
highest ccp, on which the spectral classification was based,
ranged from 0.7 to 0.95, with most being higher than 0.8.

In some cases there were two ccp (usually for adjoining spectral
types in the standard star grid) with comparable heights. For
these, we adopted the average spectral type as best representing
the spectral class of the object.

As an additional check, we inspected  the spectra visually and
compared them against those of the standard stars. In very few
cases did the spectral type yielded by the cross-correlation
method described above have to be modified, after visual
inspection. These modifications were always within our estimated
error of 0.3 of a spectral type (see below).

The adopted spectral type for each objects is reported in the last
column of Table 1.

\begin{figure}
\caption[]{Results of cross-correlation between the spectrum of
object USNO-B1 1201-0020193 with each standard star spectrum. On
the x-axis, the spectral type of the standard star is indicated,
while on the y-axis the height of the corresponding
cross-correlation peak is given. }
\end{figure}

The accuracy of the spectral classification achieved in this
manner depends on the fineness of the grid of standard spectra
used and on the signal-to-noise ratio of the cross-correlated
spectra. We estimated the accuracy of the spectral classification
in the following manner: We treated each standard star as an
object spectrum. We cross-correlated it with all of the other
standards, and determined its spectral type using the
cross-correlation peak curves, as was done for the object spectra.
In Fig. 4 we plot the derived spectral type for the standards as a
function of the reference spectral type. A least-squares linear
fit yields a slope of $0.98\pm0.06$ and a scatter of 0.3 in
spectral type.

\begin{figure}
\caption[]{Derived versus standard spectral type for the standard
stars}
\end{figure}

The effect of the S/N on the derived spectral type was estimated
by utilizing the individual exposures of various targets.
Following the same procedure as for the co-added spectra we
derived spectral types and compared them to the spectral types
adopted for the co-added spectra. Although the cross-correlation
peaks were lower for the lower S/N spectra, the maximum value of
the cross-correlation peaks was equally well defined and no
differences in spectral classification were noted down to a S/N of
about 30. There are two spectra for which the S/N is significantly
lower and for these the spectral classification is less accurate
(spectral type followed by the symbol "::", in Table 1).
Therefore, for almost all of our spectra the accuracy of the
spectral classification depends on the fineness of the
standard-star grid and is around 0.3 of a spectral type.

For the three spectra obtained in September 2003, the
cross-correlation was performed over the common wavelength range
between these spectra and the standards (that were obtained in the
subsequent observing runs), i.e. over 1000\AA. Despite the shorter
wavelength range, the cross-correlation peaks were very high,
similar to those obtained for the rest of the spectra. As
mentioned earlier, one of the stars ([PMH2004] 3 in M~33) was
observed in both wavelength ranges. The spectral classes derived
were in excellent agreement (within 0.1 of a spectral type).

\section{Discussion}
All the objects observed, except for [PMH2004] 358 in the
direction of M~33 which is a known galaxy (LEDA 5899) and which is
discussed separately below, are consistent with being foreground
stars, the spectral types of which range from A5 to M5. Figure 5
shows the frequency distribution of the derived spectral types.

In Table 2, we list some of the X-ray properties of the sources
(as given in PMH2004 and PFH2005), namely the X-ray flux, $f_x$,
the logarithm of the ratio of the X-ray flux to the optical flux,
$log(f_x/f_{opt})$ and  the hardness ratios HR1 and HR2.

The fluxes $f_x$ were calculated from the 0.2-4.5 keV XID band.
The ratios $log(f_x/f_{opt})$ were calculated using the formula
$log(f_x/f_{opt})=log(f_x)+(B+R)/(2 \times 2.5)+5.37$(Maccacaro et
al. 1988) and the USNO-b1 $B$ and $R$ magnitudes given in Table 1.
The hardness ratios were derived from the count rates in the
energy bands $R_1=0.2-0.5$ keV, $R_2=0.5-1.0$ keV, and
$R_3=1.0-2.0$ keV, according to the formulae
$HR1=(R_2-R_1)/(R_2+R_1)$ and $HR2=(R_3-R_2)/(R_3+R_2)$, as in
PMH2004.

The X-ray fluxes span two orders of magnitude, ranging from
1.5e-15 mW/m$^2$ to 1.47e-13 mW/m$^2$ . The values of
$log(f_x/f_{opt})$ range from -5.9 to -2.5, with an average of
$-4.2\pm0.8$. HR1 takes values between 0.16 and 0.94, with an
average of $0.5\pm0.2$, while HR2 ranges from -0.93 to 0.32, with
an average of $-0.5\pm0.3$. There is no clear correlation between
any of the X-ray properties listed here and the derived spectral
types of the  optical counterparts. However, the object with the
lowest X-ray to optical flux ratio is the only early type star in
the sample (of A type).

All of the objects included in the present study, except for the
known galaxy LEDA 5899, satisfy the basic criterion for being
foreground stars, i.e. $log(f_x/f_{opt})<-1$ (Maccacaro et al.
1988 ). They also satisfy the additional criterion described in
PFH2005, HR2$-$EHR2$<$0.3 (where EHR2 is the error in the value of
HR2, also given in Table 2). In Figure 6 we plot the hardness
ratio HR2 against the X-ray to optical flux ratio
[$log(f_x/f_{opt})$], marking the different spectral classes in
different colours. It is clear that all spectral classes in our
sample (with enough stars to make the comparison meaningful)
occupy the same locus on the $log(f_x/f_{opt})$-HR2  plane.
However, there are three objects, i.e. [PFH2005] 464 in the
direction of M~31 and [PMH2004] 206 and [PMH2004] 337 in the
direction of M~33,  that have hardness ratios too high compared to
the rest of the sample (although they do satisfy the previously
mentioned criterion for foreground stars). Further inspection of
these sources showed that two of them may have overestimated
hardness ratios due to their proximity to bright hard sources (see
discussion of individual objects below). Further discussion of the
three sources follows in paragraphs 5.1, 5.5 and 5.7.

\begin{figure}
\caption[]{Frequency distribution of spectral types of the
observed bright stellar counterparts of X-ray sources in M~31 and
M~33}
\end{figure}

\begin{figure}
\caption[]{X-ray properties of the observed sample. Different
spectral types are marked in different colours.}
\end{figure}

\begin{table*}
\begin{center}
\caption{{\bf X-ray properties of the objects of Table 1.}}

\begin{tabular}{lcccc}
\hline\hline\noalign{\smallskip} \noalign{\smallskip}
\multicolumn{5}{c}{{M~31}}\\
\noalign{\smallskip}\hline\noalign{\smallskip}
{ID$^1$}&{$f_x$ (mW/m$^2$)}&{{\bf$log(f_x/f_{opt})$}} &{HR1} &{HR2} \\
\noalign{\smallskip} \hline\noalign{\smallskip}
26  &8.3e-15$\pm$9e-16 & -3.6 $\pm$0.1&0.47$\pm$0.08&-0.52$\pm$0.09\\
31  &2.4e-14$\pm$1e-15 & -4.0 $\pm$0.1&0.71$\pm$0.04&-0.60$\pm$0.04\\
49  &6.9e-15$\pm$7e-16 & -4.2 $\pm$0.1&0.74$\pm$0.08&-0.48$\pm$0.07\\
59  &7.0e-15$\pm$1e-15 & -4.6 $\pm$0.2&0.39$\pm$0.16&-0.39$\pm$0.19\\
101 &6.6e-15$\pm$5e-16 & -4.1 $\pm$0.1&0.75$\pm$0.05&-0.69$\pm$0.06\\
137 &1.1e-15$\pm$4e-16 & -4.8 $\pm$0.4&0.16$\pm$0.21&-0.47$\pm$0.27\\
168 &2.9e-14$\pm$1e-15 & -3.6 $\pm$0.1&0.68$\pm$0.03&-0.58$\pm$0.04\\
217 &2.0e-15$\pm$4e-16 & -5.2 $\pm$0.2&0.50$\pm$0.12&-0.93$\pm$0.13\\
464 &1.5e-15$\pm$5e-16 & -5.1 $\pm$0.3&0.24$\pm$0.24& 0.07$\pm$0.22\\
479 &4.9e-15$\pm$5e-16 & -4.6 $\pm$0.1&0.68$\pm$0.08&-0.53$\pm$0.08\\
498 &2.9e-15$\pm$6e-16 & -4.7 $\pm$0.2&0.72$\pm$0.17&-0.60$\pm$0.16\\
553 &1.5e-14$\pm$1e-15 & -4.0 $\pm$0.1&0.58$\pm$0.07&-0.63$\pm$0.08\\
615 &2.1e-15$\pm$6e-16 & -4.4 $\pm$0.3&0.30$\pm$0.18&-0.79$\pm$0.22\\
733 &2.0e-15$\pm$5e-16 & -4.5 $\pm$0.3&0.19$\pm$0.23&-0.33$\pm$0.27\\
\noalign{\smallskip} \hline\noalign{\smallskip}
\multicolumn{5}{c}{{M~33}}\\
\noalign{\smallskip}
\hline\noalign{\smallskip}
3  &6.33e-15$\pm$1.7e-15& -4.1 $\pm$0.3&0.37$\pm$0.21&-0.64$\pm$0.26\\
28 &2.58e-14$\pm$2.1e-15& -3.0 $\pm$0.1&0.44$\pm$0.07&-0.33$\pm$0.07\\
39 &6.29e-15$\pm$8.7e-16& -2.6 $\pm$0.2&0.49$\pm$0.10&-0.24$\pm$0.11\\
54 &7.51e-15$\pm$1.4e-15& -3.9 $\pm$0.2&0.64$\pm$0.15&-0.75$\pm$0.14\\
77 &2.34e-14$\pm$3.3e-15& -3.6 $\pm$0.2&0.67$\pm$0.11&-0.48$\pm$0.12\\
92 &2.20e-15$\pm$8.2e-16& -5.2 $\pm$0.4&0.85$\pm$0.14&-0.75$\pm$0.18\\
122$^2$&4.85e-15$\pm$1.7e-15& -3.0 $\pm$0.4$^2$&0.94$\pm$0.32&-0.36$\pm$0.29\\
142&2.78e-15$\pm$7.8e-16& -5.5 $\pm$0.3&0.73$\pm$0.20&-0.51$\pm$0.22\\
182&2.94e-15$\pm$5.4e-16& -5.9 $\pm$0.2&0.35$\pm$0.13&-0.54$\pm$0.16\\
196$^3$&1.49e-14$\pm$8.6e-16&              &0.38$\pm$0.05&-0.65$\pm$0.05\\
200&1.43e-14$\pm$9.4e-16& -3.9 $\pm$0.1&0.61$\pm$0.05&-0.52$\pm$0.05\\
204&5.17e-15$\pm$6.5e-16& -4.6 $\pm$0.1&0.46$\pm$0.10&-0.73$\pm$0.11\\
206&6.05e-15$\pm$1.1e-15& -3.5 $\pm$0.2&0.61$\pm$0.32& 0.32$\pm$0.20 \\
281&2.10e-15$\pm$5.4e-16& -5.0 $\pm$0.3&0.31$\pm$0.17&-0.69$\pm$0.18\\
297&7.80e-14$\pm$2.2e-15& -2.5 $\pm$0.1&0.56$\pm$0.03&-0.10$\pm$0.03\\
337&2.09e-14$\pm$1.9e-15& -3.3 $\pm$0.1&0.33$\pm$0.12& 0.30$\pm$0.09\\
358$^4$&9.51e-15$\pm$1.6e-15&              &0.56$\pm$0.20&-0.14$\pm$0.16\\
372&2.60e-15$\pm$7.7e-16& -4.7 $\pm$0.3&0.80$\pm$0.18&-0.54$\pm$0.21\\
406&1.47e-13$\pm$5.3e-15& -3.6 $\pm$0.1&0.47$\pm$0.04&-0.41$\pm$0.03\\
\noalign{\smallskip} \hline
\end{tabular}
\end{center}

$^1$X-ray source, as in Table 1\\
$^2$ For the calculation of $log(f_x/f_{opt})$ for this object we
have used the R magnitude of the USNO-B1 counterpart given in
Table 1, which is most probably a blend of two fainter stars (see
Section 5.3). Therefore, the derived ratio $log(f_x/f_{opt})$
should be treated as a lower limit.\\
$^3$ This source was not identified in the USNO-B1 catalogue (see
section 5.4), but only in the 2-MASS catalogue. Thus, there
is no estimate of $f_{opt}$, and therefore of $log(f_x/f_{opt})$ for this object.\\
$^4$The published optical and X-ray fluxes do not necessarily
correspond to the same
contour of the galaxy, thus we did not attempt to calculate $log(f_x/f_{opt})$ for this object.\\


\end{table*}

\subsection {[PFH2005] 464, in the M~31 field (USNO-B1 1311-0013521)}
This object has a hardness ratio (HR2)  value that is high
compared to other stars in the sample with similar X-ray to
optical flux ratios (as seen in Fig.6). The optical spectrum of
the star shows no indication of emission lines that might indicate
flaring activity. However, the X-ray source itself is quite faint
and relatively close to two bright hard sources [PFH2005] 457
($f_x=3.9e-13 \pm 5e-15$, $HR2=0.46 \pm 0.01$) and [PFH2005] 463
($f_x=2.7e-13 \pm 4e-15$, $HR2=0.81 \pm 0.02$). Both sources are
at a distance of 1.52 arcmin from [PFH2005] 464 and have higher
X-ray fluxes by two orders of magnitude. This may disturb the
X-ray detection program in the determination of the hard count
rate and lead to a harder HR2.

\subsection {Source south-west of [PMH2004] 3 in the M~33 field}
This object, named 3b in Table 1, was not included in the
catalogue of X-ray sources of PMH2004. Its position on a EPIC PN
CCD boundary, about 30\arcsec\ south-west of [PMH2004] 3 and far
off-axis, prevented its detection by automatic procedures.
However, the 0.2--1 keV X-ray image (see smoothed contour overlay
in Fig. 7) clearly indicates X-ray emission, most likely
originating from the M3 star USNO-B1 1204-0019500 (R magnitude
14.66).

\begin{figure}
\caption[]{X-ray contours from the (0.2--1.0) keV combined image
of PMH2004 overlaid on the optical R image of Massey et al. The
contours are at levels 3, 4 and $6\times10^{-6}$ cts s$^{-1}$
pixel$^{-1}$, and indicate the position of additional X-ray
emission coinciding with the USNO-B1 source 1204-0019500. The
position on a pn CCD gap prevented the detection of this emission
as an X-ray source. The nearby source [PMH2004]\,3 is shown with a
box whose size corresponds to the 3 sigma error.}
\end{figure}

The spectrum we obtained for this object is that of an M star.

\subsection {[PMH2004] 122, in the M~33 field}
This source correlates with the bright USNO-B1 star 1210-0020531.
However, inspection of the deeper 4m Mosaic images of M~33 from
the Kitt Peak National Observatory archive (from the Local Group
Survey of Massey et al. 2002) shows that there are actually two
relatively bright stars separated by $\sim$6\arcsec\ in NNE/SSW
direction which are probably blended into one star in the USNO-B1
catalogue. We have obtained spectra for both stars. They both have
G-type spectra, i.e. both stars are compatible with the known
X-ray properties of the X-ray source [PMH2004] 122 (see Table 2).
Thus, it has not been possible to confirm which of the two stars
is the optical counterpart of the X-ray source. More accurate
position of the X-ray source would help resolve this question.
Unfortunately, the source is outside the FOV of the archival
\chandra\ observations.

\subsection {[PMH2004] 196, in the M~33 field (2MASS01334186+3038491)}
Due to the position of [PMH2004] 196 close to the center of M~33,
its optical counterpart was not catalogued in USNO-B1 but only in
2MASS. We therefore give the 2MASS position in Table 1. The
optical spectrum identifies it as a late M type star, which shows
significant H$_{\alpha}$ emission (see Figure 8). The emission
line is clearly seen in all four separate exposures obtained for
this star (the combined spectrum from all four exposures is shown
in Fig. 8). The equivalent width of the emission line is estimated
to be $5.7\pm0.3$ \AA. The presence of Balmer emission lines in
late M dwarfs is quite frequent and is linked to variability (e.g.
Jaschek \& Jaschek 1987).  Flare stars are expected to exhibit a
late spectral type (typically dwarf M-type) and to present
emission in CaII (not included in our spectral range) and in the
Balmer lines, during the quiescent phase. The strength of the
Balmer emission reaches a maximum during the second phase of a
flaring event (i.e. after the first strong continuum phase).

The object displays long-term X-ray variability, changing its
average flux by a factor of 6, most likely due to different
amounts of flaring activity during the observations. A long term
\xmm\ EPIC light curve is presented in Misanovic et al. (2005,
their source 170). The source is also found to be variable in the
analysis of Chandra data (Grimm et al. 2005). We analyzed
\chandra\ data from the archive to search for flaring. While the
source was faint during observation 1730, it showed strong flaring
activity during observation 786, when it was found to be brighter
on average (see Figure 9). Although the source is faint in X-rays
we managed to produce an \xmm\ EPIC X-ray spectrum, by combining
the counts from three observations. We have obtained an acceptable
fit to the spectrum for a plasma with kT of 2.8 keV.  Such a high
temperature is expected during flares.

\begin{figure}
\caption[]{Optical spectrum  (not flux normalized) of the optical
counterpart of [PMH2004]~196 in M~33, identifying it as a late M
star with H$_{\alpha}$ emission.}
\end{figure}

\begin{figure}
\caption[]{Chandra ACIS S lightcurve of [PMH2004]~196 in M~33
during observation 786 integrated over 1000~s. Time zero
corresponds to the start of the observation.}
\end{figure}

\subsection {[PMH2004] 206 in the M~33 field (USNO-B1 1209-0020969)}
This is an early-K star which has a somewhat high hardness ratio
and it was classified as a possibly hard source by PMH2004. The
X-ray flux does not show significant variability within the
errors, which are quite high due to the faintness of the source.
Moreover, the optical spectrum of the optical counterpart shows no
indication of emission lines that might indicate flaring activity.
However, the hardness ratio of the source does show some evidence
of variability between observations that are spread over three
years (Figure 10). As can be seen in this figure, the source often
appears to have  a much softer hardness ratio than that given in
Table 2 (and plotted on Fig. 6).

There seems to be at least one other early K giant that shows
significant coronal activity and some temperature variability
(Audard et al. 2004). The authors suggested that most of the
coronal heating in that star was contributed by small flares.

\begin{figure}
\caption[]{Hardness ratio (HR2 as defined in PMH2004) as a
function of epoch of observation for [PMH2004] 206, which
corresponds to source 176 in Misanovic et al. (2005). The time is
given in days, with the reference day at JD 2450814.5.}
\end{figure}

\subsection {[PMH2004] 297, in the M~33 field (USNO-B1 1209-0021478)}
In PMH2004 this object was misidentified as a background
elliptical galaxy (object MD 53 in Christian and Schommer 1982).
It is actually a stellar source to the North of this galaxy.
Indeed, the spectrum is a stellar spectrum of late G-type.

The source displays significant long-term variability in X-rays,
changing its flux by a factor of 4. The light curve is presented
in Misanovic et al. (2005, their source 253). [PMH2004] 297 is one
of the two sources among the objects studied here for which an
X-ray spectrum could be produced. Preliminary analysis of the
spectrum indicates that it is typical of very active stars (for
example see Guedel 2001, 2004). We have obtained an acceptable fit
with a combination of two temperature plasmas with kT of 0.8 and
2.3 keV. The high temperature component is most probably produced
by flaring or some other type of stellar activity, which may also
explain the significant X-ray variability of this source.

\subsection {[PMH2004] 337, in the M~33 field (USNO-B1 1204-0021198)}
This object has a hardness ratio (HR2){\footnote  {The source is
covered by three individual XMM-Newton observations and detected
above the selected threshold in two of them. The HR2 seems to be
approximately the same (0.24 and 0.28) in both observations
separated by several months.}} value that is too high compared to
other stars in the sample with similar X-ray to optical flux
ratios. The optical spectrum of the star shows no indication of
emission lines that might indicate flaring activity. However, the
X-ray source itself is relatively close (0.99 arcmin) to a very
bright hard source, [PMH2004] 335 (HR2$=0.19 \pm 0.01$), which may
disturb the count rate and hardness ratio determination, as it is
two orders of magnitude more luminous ($f_x=1.56e-12 \pm 1.6e-14$)
than [PMH2004] 337.

\subsection {[PMH2004] 358, in the M~33 field (LEDA 5899)}
This source corresponds to the spiral galaxy LEDA 5899
(CGCG502-114) which is found within the M~33 D25 ellipse and shows
a somewhat disturbed appearance. Figure 11 shows images of the
galaxy in U and V (top and bottom panel), extracted from the 4m
Mosaic images of M~33 from the Kitt Peak National Observatory
archive (from the Local Group Survey of Massey et al. 2002). The
overlayed circles show the X-ray positions of the source and the
corresponding 3 sigma error boxes from PMH2004 (red circle) and
Misanovic et al. (2005 yellow circle). The optical spectrum of the
galaxy that we obtained here is presented in Figure 12. Some
characteristic lines are marked. Our determined redshift (0.034)
agrees within the errors with the value of 0.035 given in Huchra
et al. (1999). X-ray variability could not be established for this
object, due to lack of sufficient data. The X-ray luminosity of
the galaxy in the band (0.2-4.5)keV is found to be
$2-3\times10^{40}$erg s$^{-1}$, a luminosity typical of normal
spiral galaxies (see e.g. Fabbiano et al. 1992).

\begin{figure}
\caption[]{Images of the galaxy LEDA 5899, in U (top panel) and V
(bottom panel). The circles show the X-ray positions of the source
and the corresponding 3 sigma error boxes from PMH2004 (bright
circle) and Misanovic et al. (2005 dark).}
\end{figure}

\begin{figure}
\caption[]{Optical spectrum (not flux normalised) of Galaxy LEDA
5899.}
\end{figure}

\section{Summary}
We have obtained optical spectra for the 14 brightest optical
counterparts of the \xmm\ source catalogue of the M~31 field
(PFH2005) and the 21 brightest counterparts of the \xmm\ source
catalogue of the M~33 field (PMH2004), using the 1.3m Skinakas
telescope in Crete, Greece. All of the M~31 sources and all but
one of the M~33 sources are foreground stars, of spectral types
between A5 and M5. One of the M~33 sources (lying within the D25
ellipse) is a background galaxy. The majority of the objects have
X-ray hardness ratios and X-ray to optical flux ratios (see Table
2) that are consistent with being a foreground star. One of the
stars close to the M~33 center is a late M dwarf with H$_{\alpha}$
emission, probably a flare star,
    also displaying strong X-ray variability.

\acknowledgements The authors thank T. Koutentakis and A.
Strigachev who helped with the observations at Skinakas
Observatory and A. Zezas, for his help in determining the redshift
of LEDA 5899. We thank the anonymous referee for very helpful
comments.\\

\end{document}